\documentclass[lettersize,journal]{IEEEtran}
\usepackage[Symbol]{upgreek}
\usepackage{amsmath,amsfonts}
\usepackage{algorithmic}
\usepackage{algorithm}
\usepackage{array}
\usepackage[caption=false,font=footnotesize]{subfig}
\usepackage{textcomp}
\usepackage{stfloats}
\usepackage{url}
\usepackage{verbatim}
\usepackage[export]{adjustbox}
\usepackage{graphicx}
\usepackage{cite}
\graphicspath{{img/}}
\usepackage{mathtools}
\usepackage{changepage}

\usepackage{microtype}
\usepackage{interval}
\usepackage{booktabs,makecell,tabularx}

\newcolumntype{C}[1]{>{\centering\arraybackslash}p{#1}}
\newcolumntype{L}{>{\raggedright\arraybackslash}X}
\usepackage{siunitx}
\usepackage{etoolbox}
\usepackage{xcolor}\newcommand{\rev}[1]{#1}\newcommand{\revb}[1]{#1}
\allowdisplaybreaks
\newrobustcmd{\B}{\bfseries}
\usepackage[dvipsnames]{xcolor}
\usepackage[hidelinks]{hyperref}
\hypersetup{pdftitle={A Nested Heterogeneous Multiscale Framework for Power System Dynamic Simulation},pdfauthor={Kaiyang Huang, Min Xiong, Yang Liu, Kai Sun}}
\usepackage{amsthm}
\usepackage{amssymb}
\newtheorem{theorem}{Theorem}
\newtheorem{lemma}{Lemma}
\newtheorem{proposition}{Proposition}

\newtheorem{corollary}{Corollary}
\theoremstyle{remark}

\usepackage{balance}
\newcommand{\qmap}{\mathcal Q}
\newcommand{\rmap}{\mathcal R}
\newcommand{\AK}{\mathcal A}
\newcommand{\norm}[1]{\left\lVert#1\right\rVert}

\makeatletter
\g@addto@macro\normalsize{%
  \setlength\abovedisplayskip{4pt plus 2pt minus 1pt}%
  \setlength\belowdisplayskip{4pt plus 2pt minus 1pt}%
  \setlength\abovedisplayshortskip{1pt plus 2pt}%
  \setlength\belowdisplayshortskip{3pt plus 2pt minus 1pt}}
\makeatother
\begin{document}
\title{A Nested Heterogeneous Multiscale Framework \\ for Power System Dynamic Simulation}
\author{Kaiyang~Huang,~\IEEEmembership{Member,~IEEE,}
Min~Xiong,~\IEEEmembership{Senior Member,~IEEE,}
Yang~Liu,~\IEEEmembership{Senior Member,~IEEE,}
and~Kai~Sun,~\IEEEmembership{Fellow,~IEEE}%
\thanks{This work was supported by the National Science Foundation under Grant ECCS-2329924.}%
\thanks{K. Huang is with the Department of Electrical Engineering and Computer Science, South Dakota State University, Brookings, SD 57007 USA, and was with the Department of Electrical Engineering and Computer Science, University of Tennessee, Knoxville, TN 37996 USA (e-mail: kaiyang.huang@sdstate.edu). M. Xiong is with ISO New England (e-mail: mxiong@iso-ne.com). Y. Liu is with GridUnity(e-mail: yang.liu@gridunity.com). K. Sun is with the Department of Electrical Engineering and Computer Science, University of Tennessee, Knoxville, TN 37996 USA (e-mail: kaisun@utk.edu).}}

\markboth{Submitted to IEEE Transactions on Power Systems}%
{Huang \MakeLowercase{\textit{et al.}}: A Nested Heterogeneous Multiscale Framework for Power System Dynamic Simulation}
\maketitle
\begin{abstract}
\revb{Power system dynamics involve multiple timescales, from electromagnetic transients of the network and inverter-based resources to electromechanical oscillations of synchronous generators and quasi-steady-state evolution of the operating point. Simulating these dynamics with a detailed electromagnetic transient model is computationally prohibitive, whereas existing methods reduce the model to focus on a single timescale and lose accuracy when dynamics of different timescales interact. This paper proposes a nested heterogeneous multiscale framework for highly efficient multi-timescale simulation of a power system represented by one unified detailed model. Rather than varying only the integration step, the framework adaptively shifts the simulation resolution among three nested levels so that the dynamics of interest at each stage are always captured, with the slower dynamics estimated on the fly from short detailed simulations at controlled accuracy. Theoretical analysis proves the stability of long steps during quasi-steady-state evolution and quantifies the attainable speedup. On a modified IEEE 39-bus system with frequency regulation and load restoration, the framework accurately reproduces the dynamics of all timescales and runs 15 to 18 times faster than conventional simulation of the detailed model. The accuracy and speedup of the proposed method are further verified on a 390-bus system, demonstrating its scalability to large power systems.}
\end{abstract}

\begin{IEEEkeywords}
Electromagnetic transients, electromechanical transients, heterogeneous multiscale method, inverter-based resources, multi-timescale simulation.
\end{IEEEkeywords}

\section{Introduction}
\IEEEPARstart{I}{nverter-based} resources (IBRs) introduce converter and control dynamics in the millisecond range into power systems whose synchronous generators (SGs) oscillate over seconds and whose operating point is regulated over minutes~\cite{stability}. Processes such as frequency regulation after a generation loss or sequential load restoration therefore combine electromagnetic transients at every switching action, electromechanical oscillations, and a slow quasi-steady evolution of the operating point \rev{in one continuous process. Because converter controls interact with both the network and the slower regulation, an accurate study needs all of these dynamics in one model~\cite{stability,hybrid,interfaces}.} An electromagnetic transient (EMT) model, which represents the network by instantaneous three-phase voltages and currents together with the detailed device and controller dynamics, captures all of these behaviors in one set of \rev{differential equations}. Its integration step, however, is set by the \rev{fastest dynamics, so a process of several minutes, which EMT studies rarely cover,} requires tens of millions of steps even though the fast dynamics are active only during short intervals after each disturbance.

EMT programs \rev{typically} discretize the network and devices with the trapezoidal rule at a fixed step and apply critical damping adjustment at discontinuities~\cite{dommel,lincda,nzale}. \rev{The cost of long EMT simulations is usually reduced by simplifying the model or partitioning the simulation: phasor-based models replace the fast dynamics with algebraic relations, which becomes less reliable when IBR controls interact with the network over a wide frequency range, while multiple-time-step, hybrid EMT--phasor, and envelope-based formulations assign different steps, models, or waveform representations to different parts of the system~\cite{semlyen,shu,hybrid,interfaces,strunz,mattavelli}. Variable-step EMT simulation~\cite{shenvts} adapts the step size, but in instantaneous three-phase variables even steady operation consists of 60-Hz sinusoids, which keep the step small.} \rev{The heterogeneous multiscale method (HMM) offers another route~\cite{hmmreview,engquist,weinan,ariel}. It runs the detailed model only over short time windows, averages the rate of change of the slow variables over each window, called the effective force, and uses this rate to take steps far longer than the detailed model allows. Projective integration follows a related idea and has been nested for problems with several timescales~\cite{gear}.}

The two-level HMM in~\cite{huanghmm} brought this approach to power system simulation. \rev{It maps the EMT state to slowly varying quantities that no longer oscillate at the fundamental frequency in steady operation. The rates of these quantities, estimated from short EMT runs, advance them with steps of tens of milliseconds, and the result is mapped back to an EMT state to start the next run. These steps still follow the electromechanical oscillations, so over a process of several minutes the EMT runs are repeated thousands of times. A third level that steps through the quasi-steady evolution can remove most of this repetition. Its steps last tens of seconds, so they have to remain accurate and stable while the detailed model still contains fast dynamics. The levels also have to hand over to one another whenever a switching event occurs.}

\rev{This paper develops a three-level HMM that meets these requirements while integrating one EMT model throughout with an unmodified fixed-step EMT solver. The proposed nested HMM links the electromagnetic, electromechanical, and quasi-steady dynamics in a unified framework and recovers the complete EMT state at each level, so detailed EMT simulation resumes at any switching event. The quasi-steady level estimates its rate at the end of the averaging interval and advances with an exponential integrator that keeps residual fast modes bounded for any step length and reproduces steady operation exactly. The switching between levels follows from events and from error indicators measured against one set of accuracy requirements. The error and cost analysis shows that, at a prescribed accuracy, the quasi-steady level multiplies the speedup of the two-level HMM by a factor that grows with the separation between the electromechanical and quasi-steady timescales, and a sparse implementation extends the method to large systems. The method is validated against full EMT simulation and two-level HMM with fourth-order Runge--Kutta and nodal trapezoidal solvers on a modified IEEE 39-bus system with a restoration and nine source outages and on a 390-bus system.}

\section{EMT model and two-level HMM}\label{sec:background}
\subsection{Multi-timescale EMT model}\label{sec:model}
An EMT model of a power system is written as the piecewise-smooth ordinary differential equations (ODEs)
\begin{equation}\dot{\mathbf{x}}=\mathbf{f}_\sigma(\mathbf{x},t),\qquad \mathbf{x}(t_0)=\mathbf{x}_0,\label{eq:emt}\end{equation}
where $\mathbf{x}\in\mathbb R^{N_x}$ collects the instantaneous voltages and currents of the capacitive and inductive network elements together with the SG, IBR, and controller states, and $\sigma$ identifies the network and control configuration, which changes at switching events. \rev{Network transients and converter control loops evolve within milliseconds. Rotor angles, speeds, and primary controls evolve within seconds, and secondary regulation and load restoration evolve over minutes.} With coordinates $\mathbf{z}_\mu$, $\mathbf{z}_{M}$, and $\mathbf{z}_{Q}$ associated with the EMT, electromechanical, and quasi-steady dynamics, \eqref{eq:emt} takes the scale-separated form
\begin{equation}\epsilon_\mu\dot{\mathbf{z}}_\mu=\mathbf{f}_\mu(\mathbf{z},t),\quad\epsilon_{M}\dot{\mathbf{z}}_{M}=\mathbf{f}_{M}(\mathbf{z},t),\quad\dot{\mathbf{z}}_{Q}=\mathbf{f}_{Q}(\mathbf{z},t),\label{eq:scales}\end{equation}
with $0<\epsilon_\mu\ll\epsilon_{M}\ll1$ after time is normalized to the slowest scale. In a power system with IBRs, the network, converter, and machine states are coupled across all three groups, so a separation of the form~\eqref{eq:scales} is difficult to construct explicitly and changes with the operating condition. The HMM developed here integrates~\eqref{eq:emt} directly and uses~\eqref{eq:scales} only to organize the levels and to express the averaging error through $\epsilon_\mu$ and $\epsilon_{M}$.

\subsection{Compression, reconstruction, and effective force}\label{sec:hmm}
Following~\cite{huanghmm}, a compression map $\qmap_t$ and a reconstruction map $\rmap_t$ relate $\mathbf{x}$ to a transformed state $\mathbf{u}\in\mathbb R^{N_x}$ by $\mathbf{u}=\qmap_t(\mathbf{x})$ and $\mathbf{x}=\rmap_t(\mathbf{u})$, with $\rmap_t\circ\qmap_t=\qmap_t\circ\rmap_t=\mathbf{I}$. Here $\qmap_t$ applies the Park transformation to each three-phase group of network and interface variables and leaves the remaining states unchanged. Network variables and load currents use the rotor angle of a reference SG of their area, a tie-line current uses that of its sending area, and each IBR terminal current uses the angle of the IBR phase-locked loop (PLL). Because these angles are states of~\eqref{eq:emt}, the map is exactly invertible, and in balanced steady operation the zero-, direct-, and quadrature-axis components are constant while the angles increase linearly. The rate of $\mathbf{u}$ follows from~\eqref{eq:emt} by the chain rule,
\begin{equation}\mathbf{g}(\mathbf{u},t)=\partial_t\qmap_t(\mathbf{x})+D_{\mathbf{x}}\qmap_t(\mathbf{x})\,\mathbf{f}_\sigma(\mathbf{x},t),\qquad\mathbf{x}=\rmap_t(\mathbf{u}),\label{eq:transformed}\end{equation}
\rev{where $D_{\mathbf{x}}\qmap_t$ is the Jacobian of $\qmap_t$ with respect to $\mathbf{x}$,} so evaluating $\mathbf{g}$ amounts to reconstructing the EMT state, evaluating~\eqref{eq:emt}, and transforming the result, and the transformed dynamics $\dot{\mathbf{u}}=\mathbf{g}(\mathbf{u},t)$ inherit every device model, limit, and control mode of~\eqref{eq:emt}.

Along a trajectory of $\mathbf{u}$, the effective force is the rate that remains after the fast dynamics are averaged out. \rev{HMM estimates it by convolving a kernel with the recent trajectory, called the history~\cite{engquist,huanghmm}.} For a history $\mathbf{u}$ on $[t-\eta,t]$, let $s\in[-\eta,0]$ denote the time relative to its endpoint, and let $K_\eta(s)=\eta^{-1}K(s/\eta)$ be a kernel with unit integral and zero endpoint values. The estimated force is
\begin{equation}\mathbf{F}(t)=\int_{-\eta}^0 K_\eta(s)\,\dot{\mathbf{u}}(t+s)\,ds=-\int_{-\eta}^0 K'_\eta(s)\,\mathbf{u}(t+s)\,ds,\label{eq:conv}\end{equation}
where the second form follows from integration by parts. The kernel returns the slope of a linear history, \rev{attenuates oscillations with periods shorter than its length,} and in the second form needs only state samples.

\subsection{Two-level HMM}\label{sec:twolevel}
\rev{The two-level HMM of~\cite{huanghmm} alternates short EMT simulations, the microscopic ($\mu$) process, with the steps of a middle ($M$) level.} A $\mu/M$ cycle starts from a reconstructed EMT state, integrates~\eqref{eq:emt} with steps $h_\mu$ over a window $\eta_\mu$, and maps the samples to a history $\mathbf{u}_\mu$ on $[t-\eta_\mu,t]$. A diagonal mask $\mathbf{P}_{M}$ selects the network and converter components that oscillate within this window, and with $K_{M}$ scaled to $[-\eta_\mu,0]$ and $\mathbf{u}=\mathbf{u}_\mu(t)$, the middle force and the middle advance are
\begin{gather}\mathbf{F}_{M}=-\mathbf{P}_{M}\int_{-\eta_\mu}^0 K'_{M}(s)\,\mathbf{u}_\mu(t+s)\,ds+(\mathbf{I}-\mathbf{P}_{M})\,\mathbf{g}(\mathbf{u},t),\label{eq:mforce}\\
\mathbf{u}^+=\mathbf{u}+h_{M}\mathbf{F}_{M},\qquad \mathbf{x}^+=\rmap_{t+h_{M}}(\mathbf{u}^+),\label{eq:madvance}\end{gather}
and $\mathbf{x}^+$ starts the next microscopic window at $t+h_{M}$. The forward Euler step~\eqref{eq:madvance} is the middle macro-solver, with $h_{M}$ set by the controller of~\cite{huanghmm} under a tolerance and a cap. One cycle covers $\Delta_{M}=\eta_\mu+h_{M}$ of simulated time with $\eta_\mu/h_\mu$ microscopic steps, so the speedup over full EMT simulation is close to $1+h_{M}/\eta_\mu$\rev{, which remains modest because $h_{M}$ must resolve the electromechanical dynamics}.

\section{Proposed three-level HMM}\label{sec:method}
\subsection{Nested levels on one EMT model}\label{sec:nested}
\begin{figure}[!ht]
\color{black}
\centering\includegraphics[width=\columnwidth]{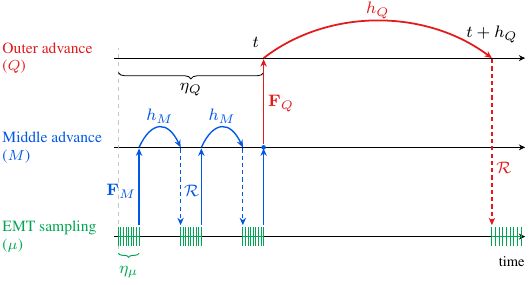}
\caption{Three-level HMM on one EMT model. Upward arrows supply effective forces, and downward arrows reconstruct the EMT state.}\label{fig:hierarchy}
\end{figure}

During quasi-steady evolution, the trajectory generated by repeated $\mu/M$ cycles \rev{varies slowly except for residual electromechanical and electrical components, so HMM can be applied once more.} The three-level HMM in Fig.~\ref{fig:hierarchy} therefore treats the middle-level trajectory as the microscopic process of an outer quasi-steady ($Q$) level. Repeated $\mu/M$ cycles over a duration $\eta_{Q}$ produce state samples whose interpolation forms the middle-level history $\mathbf{u}_{M}$ ending at $t$, with $\mathbf{u}=\mathbf{u}_{M}(t)$. A second convolution with a kernel $K_{Q}$ supported on $[-\eta_{Q},0]$ estimates the outer force
\begin{equation}\mathbf{F}_{Q}=-\int_{-\eta_{Q}}^0 K'_{Q}(s)\,\mathbf{u}_{M}(t+s)\,ds,\label{eq:qforce}\end{equation}
which acts on every component of $\mathbf{u}$, including the rotation angles. The outer macro-solver of Section~\ref{sec:outer} uses $\mathbf{F}_{Q}$ to advance the state by $h_{Q}$, the result is reconstructed as $\mathbf{x}^+=\rmap_{t+h_{Q}}(\mathbf{u}^+)$, and fresh $\mu/M$ cycles collect the next history, so a complete outer cycle covers $\Delta_{Q}=\eta_{Q}+h_{Q}$ and the full EMT state is available at the end of every advance. The levels use the microscopic solver only through state-to-state advances, so any EMT solver that advances $\mathbf{x}$ from a given state, such as an explicit Runge--Kutta method, the nodal trapezoidal discretization of EMT programs~\cite{dommel,lincda}, \rev{or variable-step and hybrid EMT--phasor solvers~\cite{shenvts,hybrid},} can serve as the $\mu$ process without modification.

\subsection{Endpoint-aligned kernel}\label{sec:kernel}
The outer advance starts at the end of the history, so $\mathbf{F}_{Q}$ must describe the rate at $t$. A kernel that is symmetric on $[-\eta_{Q},0]$ reproduces the rate at the center of the history, which differs from the rate at $t$ whenever the frequency or a control command changes over the history. The outer kernel is therefore built from shifted copies of a smooth averaging kernel, combined so that their lags cancel. Let $b_w$ be a smooth, normalized kernel supported on $[-w,0]$ whose value and derivatives vanish at both ends, with $K_{M}=b_{\eta_\mu}$ at the middle level. For $0<w<\eta_{Q}$ and $p$ distinct shifts $a_j\in[-\eta_{Q}+w,0]$, with $w$ and $a_j$ proportional to $\eta_{Q}$, the outer kernel is
\begin{equation}K_{Q}^{(p)}(s)=\sum_{j=1}^p\alpha_j\, b_w(s-a_j).\label{eq:endkernel}\end{equation}
Expanding a smooth rate about $t$ shows that the estimate reproduces its value at $t$ when the kernel preserves the constant term and cancels the first $p-1$ powers of the offset,
\begin{equation}\int_{-\eta_{Q}}^0s^m K_{Q}^{(p)}(s)\,ds=\delta_{m0},\qquad m=0,\ldots,p-1,\label{eq:moments}\end{equation}
where $\delta_{m0}$ is one for $m=0$ and zero otherwise. The moments of the shifted kernels form a lower-triangular transformation, with unit diagonal, of the Vandermonde matrix of the shifts, so for distinct shifts~\eqref{eq:moments} determines $\alpha_j$ uniquely, and the discrete weights of~\eqref{eq:qforce}, which multiply state samples, satisfy the equivalent conditions obtained from~\eqref{eq:moments} by integration by parts. Below, $K_{Q}$ denotes $K_{Q}^{(p)}$ for the order $p$ in use. The difference between the estimates of orders $p$ and $p-1$ measures the remaining lag and enters the step control of Section~\ref{sec:control}.

\subsection{Outer macro-solver}\label{sec:outer}
The outer step can exceed the history by two orders of magnitude, so the macro-solver has to keep the residual fast content of $\mathbf{F}_{Q}$ from growing with $h_{Q}$ while carrying the slow content accurately. \rev{The Jacobian of the transformed model distinguishes the two kinds of content.} At the endpoint, let $\mathbf{J}=D_{\mathbf{u}}\mathbf{g}(\mathbf{u},t)$, whose evaluation is given in Appendix~\ref{app:impl}. In directions associated with strongly decaying modes, the instantaneous field $\mathbf{g}$ already describes the relaxed evolution, whereas in slowly varying directions the convolution carries the averaged rate. \rev{The resolvent filter $\mathbf{S}=(\mathbf{I}-\tau\mathbf{J})^{-2}$, $\tau>0$, separates the two. It is defined whenever $1/\tau$ is not an eigenvalue of $\mathbf{J}$,} and the outer force passed to the propagator is
\begin{equation}\widetilde{\mathbf{F}}_{Q}=\mathbf{g}(\mathbf{u},t)+\mathbf{S}\bigl(\mathbf{F}_{Q}-\mathbf{g}(\mathbf{u},t)\bigr).\label{eq:resolvent}\end{equation}
For a mode with eigenvalue $\lambda$, the filter scales the difference between the two rates by $(1-\tau\lambda)^{-2}$, which approaches one for slow modes and decays as $|\lambda|^{-2}$ for fast ones.

The filtered force is propagated with the linearized dynamics, which the macro-solver integrates exactly. For an autonomous transformed field, the frozen increment $\mathbf{d}(\xi)$ satisfies $\dot{\mathbf{d}}=\mathbf{J}\mathbf{d}+\widetilde{\mathbf{F}}_{Q}$ with $\mathbf{d}(0)=\mathbf{0}$, and a nonlinear correction completes the step,
\begin{align}
\mathbf{d}&=h_{Q}\,\varphi_1(h_{Q}\mathbf{J})\,\widetilde{\mathbf{F}}_{Q},\label{eq:stage}\\
\mathbf{N}(\mathbf{d})&=\mathbf{g}(\mathbf{u}+\mathbf{d},t)-\mathbf{g}(\mathbf{u},t)-\mathbf{J}\mathbf{d},\label{eq:remainder}\\
\mathbf{u}^+&=\mathbf{u}+\mathbf{d}+2h_{Q}\,\varphi_3(h_{Q}\mathbf{J})\,\mathbf{N}(\mathbf{d}),\label{eq:qstep}
\end{align}
where \rev{the functions} $\varphi_j(z)=\int_0^1e^{(1-\xi)z}\xi^{j-1}/(j-1)!\,d\xi$, \rev{with} $\varphi_1(z)=(e^z-1)/z$, \rev{are evaluated at $h_{Q}\mathbf{J}$ as matrix functions,} and $\mathbf{N}$ measures the departure of the transformed field from its linearization. For $\widetilde{\mathbf{F}}_{Q}=\mathbf{g}$, \eqref{eq:stage}--\eqref{eq:qstep} is the third-order exponential Rosenbrock method~\cite{hochbruck,hochbruckacta}, and the outer macro-solver extends it by supplying the HMM force through~\eqref{eq:resolvent}. Explicit time dependence is handled by appending time to the state with unit rate. Equations~\eqref{eq:qforce}--\eqref{eq:qstep} together define one outer cycle, whose stability and accuracy are analyzed in Section~\ref{sec:analysis}.

\subsection{Error scales, step control, and level transitions}\label{sec:control}
\rev{Accuracy is specified by component scales $w_i>0$ in the norm $\norm{\mathbf{v}}_{\mathbf{W}}=\max_i|v_i|/w_i$, each reflecting the accuracy required of the quantity that the state drives.} A state that acts on another quantity through a gain inherits that quantity's scale divided by the gain. A common factor $\sigma_{Q}$ multiplies all outer scales and sets the requested accuracy.

The outer step is selected from two error indicators, the size of the nonlinear correction in~\eqref{eq:qstep} and the change in the propagated increment when the kernel order is reduced from $p$ to $p-1$,
\begin{align}
e_{\mathrm{nl}}&=\norm{2h_{Q}\varphi_3(h_{Q}\mathbf{J})\mathbf{N}(\mathbf{d})}_{\mathbf{W}},\label{eq:indicators}\\
e_{\mathrm{k}}&=\norm{h_{Q}\varphi_1(h_{Q}\mathbf{J})\,\mathbf{S}\,\bigl(\mathbf{F}_{Q}^{(p)}-\mathbf{F}_{Q}^{(p-1)}\bigr)}_{\mathbf{W}},\label{eq:kernelindicator}
\end{align}
and a trial step is accepted when $\max(e_{\mathrm{nl}},e_{\mathrm{k}})\le1$. \revb{Section~\ref{sec:restoration} evaluates the actual local error of every accepted step.} For fixed sampling data and a smooth field, $\mathbf{d}=O(h_{Q})$ and $\mathbf{N}(\mathbf{d})=O(h_{Q}^2)$, so $e_{\mathrm{nl}}$ scales as $h_{Q}^3$ and $e_{\mathrm{k}}$ as $h_{Q}$, which gives the controller
\begin{equation}h_{Q}^{\mathrm{new}}=\gamma\, h_{Q}\min\bigl\{e_{\mathrm{nl}}^{-1/3},\,e_{\mathrm{k}}^{-1}\bigr\}\label{eq:controller}\end{equation}
with a safety factor $\gamma<1$ and limits on the step and its growth. A rejected trial is repeated with the reduced step, reusing the history, the forces, and $\mathbf{J}$, and a matrix-function approximation that fails to converge counts as a rejected trial.

The same scales decide which level advances the solution. A switching event applies its state reset and starts detailed EMT integration, which proceeds in windows of length $\eta_\mu$. After each window, the transition indicator $e_\mu=\max_{i\in\mathcal F}\max_j|u_i(t_j)-\hat u_i(t_j)|/w_i$ compares the samples $u_i(t_j)$ of the network, interface, and load components $\mathcal F$ with their least-squares line $\hat u_i$ over the window, with the scales evaluated at $\sigma_{Q}=1$. Once $e_\mu\le1$ in two consecutive windows, the remaining electrical transient lies below the error scales and $\mu/M$ cycles take over. After every middle-level history of length $\eta_{Q}$, an outer step is attempted, and a step that the controller limits to less than $h_{Q,\min}=2\eta_{Q}$ is deferred while the $\mu/M$ cycles collect the next history, because a shorter step saves little relative to \rev{the cost of} the history and the Jacobian. \rev{Through this rule, $e_{\mathrm{k}}$ also governs the hand-over to the outer level, since residual electromechanical components in the history separate the estimates of orders $p$ and $p-1$.} Sampling and advancement stop at the next event, supervisory check, or final time, so no level crosses an event, and a check that leaves the configuration unchanged continues directly with the next sampling cycle. The length of every detailed and middle interval thus follows from the simulated response, as Algorithm~\ref{alg:hmm} summarizes.

\begin{algorithm}[!t]
\caption{Three-Level HMM}\label{alg:hmm}
\begin{algorithmic}[1]
\renewcommand{\algorithmicrequire}{\textbf{Input:}}
\REQUIRE EMT model and solver, initial state, final time $T$, settings, and event and supervisory rules.
\STATE Initialize $t$, $\mathbf{x}$, $\mathbf{u}$, $h_{Q}$, and the level $\ell\leftarrow$ detailed.
\WHILE{$t<T$}
\STATE If an event or check is due, process it. On a configuration change, reset the state, set $\ell\leftarrow$ detailed, and clear the history.
\STATE Bound all advances by the next event, check, or $T$.
\IF{$\ell=$ detailed}
\STATE Integrate~\eqref{eq:emt} over $\eta_\mu$ and evaluate $e_\mu$. Set $\ell\leftarrow$ middle after two windows with $e_\mu\le1$.
\ELSIF{the history is shorter than $\eta_{Q}$}
\STATE Run a $\mu/M$ cycle by~\eqref{eq:mforce}--\eqref{eq:madvance} and add it to the history.
\ELSE
\STATE Evaluate~\eqref{eq:qforce}, $\mathbf{g}$, $\mathbf{J}$, and~\eqref{eq:resolvent}.
\REPEAT
\STATE Form~\eqref{eq:stage}--\eqref{eq:qstep}, evaluate~\eqref{eq:indicators}--\eqref{eq:kernelindicator}, and update $h_{Q}$ by~\eqref{eq:controller}.
\UNTIL{$\max(e_{\mathrm{nl}},e_{\mathrm{k}})\le1$ or $h_{Q}<h_{Q,\min}$}
\STATE If accepted, set $\mathbf{u}\leftarrow\mathbf{u}^+$, advance $t$, and reconstruct $\mathbf{x}=\rmap_t(\mathbf{u})$. Otherwise defer. Clear the history.
\ENDIF
\ENDWHILE
\end{algorithmic}
\end{algorithm}

\section{Error, stability, and cost analysis}\label{sec:analysis}
Errors in the three-level HMM move upward through the levels. Microscopic integration errors enter the middle-level history, the history error and the averaging bias enter the outer force, and the outer macro-solver carries the endpoint and force errors into the next cycle. The reference is the ideal effective dynamics of~\eqref{eq:scales}. Let $\mathbf{F}_{M}^\star(\mathbf{u},t)$ and $\mathbf{F}_{Q}^\star(\mathbf{u},t)$ be the ideal effective forces at the middle and outer levels, in which the dynamics faster than $\epsilon_\mu$ and $\epsilon_{M}$, respectively, are averaged out~\cite{hmmreview}, with the ideal effective state $\mathbf{u}^\star$ following $\mathbf{g}$ during detailed integration, $\mathbf{F}_{M}^\star$ during $\mu/M$ cycles, and $\mathbf{F}_{Q}^\star$ during outer steps. The exact history of a micro window is the flow of $\mathbf{g}$, and that of the outer level is the flow of $\mathbf{F}_{M}^\star$, each from the state at its start. Reference quantities at the same physical times carry a superscript $\star$, and estimates hold on intervals with fixed configuration and limiter active set, on which $\qmap_t$ and $\rmap_t$ are smooth with bounded derivatives. \revb{Accordingly, the analysis applies to each smooth interval between changes of configuration or limiter active set, and an outer step that crosses an internally generated limiter transition without resolving its switching time lies outside it. Although such steps are not addressed by the analysis, the case study in Section~\ref{sec:restoration} demonstrates that they also achieve satisfactory accuracy.} Norms are a fixed scaled vector norm and its induced matrix norm. The proofs are given in Appendix~\ref{app:proofs}.

\subsection{Kernel estimation}\label{sec:kernelest}
Write $\AK_{K,\eta}[\mathbf{u}]$ for the convolution estimate~\eqref{eq:conv} and $\kappa_K=\int_{-1}^0|K'(s)|\,ds$ for the derivative norm of the normalized kernel, with $\kappa_{M}$ and $\kappa_{Q}$ for $K_{M}$ and $K_{Q}$.
\begin{lemma}[History-to-force gain]\label{lem:transfer}
If two histories satisfy $\sup_{s\in[-\eta,0]}\norm{\mathbf{u}(t+s)-\mathbf{u}^{\star}(t+s)}\le E$, then $\norm{\AK_{K,\eta}[\mathbf{u}]-\AK_{K,\eta}[\mathbf{u}^{\star}]}\le\kappa_KE/\eta$.
\end{lemma}

\begin{lemma}[Endpoint bias]\label{lem:bias}
For a $p$-times continuously differentiable rate $\mathbf{v}$ and a kernel $K_{Q}^{(p)}$ on $[-\eta_{Q},0]$ satisfying~\eqref{eq:moments}, the endpoint bias $\beta_p(t)=\norm{\int_{-\eta_{Q}}^0K_{Q}^{(p)}(s)\mathbf{v}(t+s)\,ds-\mathbf{v}(t)}$ satisfies
\begin{equation}\beta_p(t)\le\frac{1}{p!}\int_{-\eta_{Q}}^0|s|^p|K_{Q}^{(p)}(s)|\,ds\max_{s\in[-\eta_{Q},0]}\norm{\mathbf{v}^{(p)}(t+s)}.\label{eq:bias}\end{equation}
\end{lemma}
For the averaging error, let $B_{M}$ and $B_{Q}$ denote the norms of the differences between the forces~\eqref{eq:mforce} and~\eqref{eq:qforce} evaluated on exact histories and the ideal effective forces at their endpoints, with windows $\eta_{M}=\eta_\mu$ and $\eta_{Q}$ and fast scales $\epsilon_{f,M}=\epsilon_\mu$ and $\epsilon_{f,Q}=\epsilon_{M}$ from~\eqref{eq:scales}. If the slow force has $p_\ell$ bounded derivatives and the zero-mean fast force has $k_\ell$ uniformly bounded primitives in fast time, a kernel satisfying~\eqref{eq:moments} on its window with order $p_\ell$, whose first $k_\ell-1$ derivatives vanish at both ends and whose $k_\ell$th derivative is integrable, gives~\cite{engquist}, with a constant $C$,
\begin{equation}B_\ell\le C\bigl[\eta_\ell^{p_\ell}+(\epsilon_{f,\ell}/\eta_\ell)^{k_\ell}\bigr],\qquad\ell=M,Q.\label{eq:averaging}\end{equation}
The first term is the slow bias quantified in Lemma~\ref{lem:bias}, with $p_{M}=1$ for the single bump $K_{M}=b_{\eta_\mu}$, and the second is the incomplete cancellation of oscillations faster than the window. For an attracting transient, the second term is bounded instead by $\norm{K_\ell}_{L^1}$ times the largest amplitude of the transient force over the window.

\subsection{Stability of the outer cycle}\label{sec:stability}
\begin{figure}[!t]
\color{black}
\centering\includegraphics[width=\columnwidth]{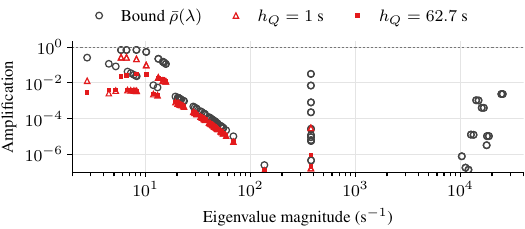}
\caption{Complete-cycle amplification of the outer macro-solver for the decaying modes \rev{at the start of the longest outer step of the restoration study (437.3~s)}.}\label{fig:stability}
\end{figure}

The outer macro-solver acts on residual fast components through the complete cycle of sampling and advancement. Its behavior is characterized on the linear test equation $\dot u=\lambda u$ with $\operatorname{Re}\lambda<0$, which describes each mode of a frozen, diagonalizable Jacobian. For a unit initial state, the difference between the kernel-weighted history and its endpoint value is $D_K(\lambda)=\int_{-\eta_{Q}}^0K_{Q}(s)e^{\lambda(\eta_{Q}+s)}ds-e^{\lambda\eta_{Q}}$.
\begin{proposition}[Step-uniform amplification]\label{prop:stability}
For $\dot u=\lambda u$ with $\operatorname{Re}\lambda<0$, exact sampling over $[0,\eta_{Q}]$ followed by~\eqref{eq:qforce}--\eqref{eq:qstep} maps the initial state $u_0$ to $R_{Q}(\lambda)u_0$ with
\begin{equation}R_{Q}(\lambda)=e^{\lambda(\eta_{Q}+h_{Q})}+\frac{e^{\lambda h_{Q}}-1}{(1-\tau\lambda)^2}\,D_K(\lambda),\label{eq:amplification}\end{equation}
and for every $h_{Q}\ge0$,
\begin{align}|R_{Q}(\lambda)|\le\bar\rho(\lambda)&=e^{\eta_{Q}\operatorname{Re}\lambda}+\frac{2|D_K(\lambda)|}{|1-\tau\lambda|^2}\nonumber\\&\le e^{\eta_{Q}\operatorname{Re}\lambda}+\frac{2(1+\norm{K_{Q}}_{L^1})}{|1-\tau\lambda|^2}.\label{eq:rhobar}\end{align}
\end{proposition}
The outer cycle therefore amplifies a decaying mode by at most $\bar\rho(\lambda)$ \rev{for any outer step length}, and the mode contracts over every cycle when $\bar\rho(\lambda)<1$, which the resolvent filter secures for strongly damped modes by reducing the second term of~\eqref{eq:rhobar} as $|\lambda|^{-2}$. Fig.~\ref{fig:stability} evaluates the proposition for the frozen 430-state Jacobian \rev{at the start of the longest outer step of the restoration study in Section~\ref{sec:restoration}, at 437.3~s,} with the settings of Table~\ref{tab:settings}. For all 378 decaying modes with $|\lambda|\ge1/\eta_{Q}$, $\bar\rho(\lambda)$ is at most 0.69, so all of them contract, and the actual amplification \rev{over this 62.7-s step} is at most 0.031.

\begin{proposition}[Force sensitivity and steady rotation]\label{prop:sensitivity}
For a diagonalizable $\mathbf{J}$ and a linear field, a perturbation of $\mathbf{F}_{Q}$ in the direction of a mode with eigenvalue $\lambda$, $\operatorname{Re}\lambda<0$, changes $\mathbf{u}^+$ by that perturbation times
\begin{equation}G(\lambda,h_{Q})=\frac{e^{\lambda h_{Q}}-1}{\lambda(1-\tau\lambda)^2},\qquad|G|\le\frac{\min\{h_{Q},\,|\operatorname{Re}\lambda|^{-1}\}}{|1-\tau\lambda|^2}.\label{eq:gain}\end{equation}
For a general field, if $\mathbf{u}(t)=\mathbf{u}_*+t\mathbf{c}$ solves $\dot{\mathbf{u}}=\mathbf{g}(\mathbf{u})$ and the history is exact, then $\mathbf{u}^+=\mathbf{u}+h_{Q}\mathbf{c}$ for every $h_{Q}$.
\end{proposition}
Force errors in slowly varying directions are thus integrated over the step, as in any one-step method, whereas in decaying directions their effect is bounded independently of $h_{Q}$ and reduced by $|1-\tau\lambda|^{-2}$. Balanced steady operation, with constant Park components and uniformly rotating angles, is advanced exactly for any step, so long outer steps add no phase drift beyond the errors of the lower-level history.

\subsection{Error estimation}\label{sec:errors}
The macro-solver of each level is a one-step method for the ideal effective dynamics, perturbed by the error of the estimated force~\cite{weinan,engquist}, which is bounded through the kernel and, at the outer level, contains the errors of the middle level. Throughout, $C$ denotes constants, possibly different at each occurrence, that depend on $T$, on the model and its switching events, and on the kernels, but not on the steps and windows. The forces $\mathbf{F}_{M}^\star$, $\mathbf{F}_{Q}^\star$, and $(\mathbf{I}-\mathbf{P}_{M})\mathbf{g}$, the flow of $\mathbf{g}$ over the detailed intervals, and the state-reset maps are Lipschitz with constant $C$. Three assumptions are made. First, the micro-solver has order $r$, so a window of length $\eta_\mu$ from an exact state has error at most $C\eta_\mu h_\mu^r$. \rev{Second, the macro-solvers are stable and consistent for the effective dynamics. The middle step~\eqref{eq:madvance} has order one for $\mathbf{F}_{M}^\star$, and with $\widetilde{\mathbf{F}}_{Q}=\mathbf{F}_{Q}^\star$, \eqref{eq:stage}--\eqref{eq:qstep} has order three for $\mathbf{F}_{Q}^\star$ when $\mathbf{q}=\mathbf{g}-\mathbf{F}_{Q}^\star$ is constant along the effective trajectories, which then solve $\dot{\mathbf{u}}=\mathbf{g}(\mathbf{u},t)-\mathbf{q}$, so that \eqref{eq:stage}--\eqref{eq:qstep} is the exponential Rosenbrock method~\cite{hochbruck} for this field. A varying $\mathbf{q}$ adds the local error $h_{Q}^2\varphi_2(h_{Q}\mathbf{J})\dot{\mathbf{q}}+h_{Q}^3\varphi_3(h_{Q}\mathbf{J})\ddot{\mathbf{q}}$ to leading order, which enters the term $\varrho$ of Theorem~\ref{thm:nested}. Either step, with step size $h$, depends on its starting state with Lipschitz constant $C$ and on the supplied force with Lipschitz constant $Ch$,} and~\eqref{eq:qstep} depends on $\mathbf{d}$ with Lipschitz constant $C$, which for the outer step~\eqref{eq:resolvent}--\eqref{eq:qstep}, with $\mathbf{g}$, $\mathbf{J}$, and $\mathbf{S}$ evaluated at the starting state and $\mathbf{F}_{Q}$ supplied, holds when $\norm{e^{s\mathbf{J}}}\le C$ for $s\ge0$, $\mathbf{J}$ is Lipschitz in $\mathbf{u}$, and $\mathbf{g}$ and $\mathbf{F}_{Q}$ are bounded, uniformly over the outer endpoints. Third, \eqref{eq:averaging} holds at both levels, with $p_{Q}=p$ and $k_{Q}=k$, for exact histories started from the computed states, including the components that~\eqref{eq:mforce} takes directly from $\mathbf{g}$.

\begin{theorem}[Convergence]\label{thm:nested}
Under these assumptions, for numerical and reference trajectories with a common initial state and a common sequence of switching times, configurations, and limiter active sets on $[0,T]$, the computed state $\mathbf{u}(T)$ satisfies
\begin{equation}\begin{split}\norm{\mathbf{u}(T)-\mathbf{u}^\star(T)}\le C\bigl[&h_{Q}^3+\eta_{Q}^{p}+(\epsilon_{M}/\eta_{Q})^{k}\\&+h_{M}+\eta_\mu^{p_{M}}+(\epsilon_\mu/\eta_\mu)^{k_{M}}+h_\mu^r+\varrho\bigr],\end{split}\label{eq:global}\end{equation}
where steps denote their maxima over the cycles, the windows $\eta_\mu$ and $\eta_{Q}$ are fixed, and $\varrho$ bounds, per unit simulated time, the implementation errors and the three defects specified below.
\end{theorem}

\revb{Theorem~\ref{thm:nested} guarantees the accuracy of outer steps within each smooth interval of the process. }The first three terms are the error of the outer level, with $\epsilon_{M}$ as the fast scale. The terms from $h_{M}$ to $h_\mu^r$ are the error of the middle level. They carry no factor of the outer window or step, because the history error grows in proportion to $\eta_{Q}$ while Lemma~\ref{lem:transfer} divides it by $\eta_{Q}$, so they set a floor below which a smaller $\sigma_{Q}$ no longer reduces the error. The implementation errors in $\varrho$ are the quadrature, Jacobian, and matrix-function errors, the interpolation error of the outer history divided by $\eta_{Q}$, and the start-up error of the micro-solver over a window divided by $\eta_\mu$. The first defect is the additional local error of the outer step. The second is the filter defect $\norm{h_{Q}\varphi_1(h_{Q}\mathbf{J})(\mathbf{I}-\mathbf{S})(\mathbf{g}-\mathbf{F}_{Q}^\star)}$ at the exact-history endpoints. It vanishes on the steady trajectory of Proposition~\ref{prop:sensitivity}, and in a decaying mode with eigenvalue $\lambda$ its factor $(e^{\lambda h_{Q}}-1)[1-(1-\tau\lambda)^{-2}]/\lambda$ has modulus at most $\min\{2h_{Q},4\tau\}$, so it is bounded independently of $h_{Q}$. The third is the window drift, the largest distance between the flows of $\mathbf{g}$ and $\mathbf{F}_{M}^\star$ over a micro window from its initial state divided by $\Delta_{M}$. The scale-separation error $\varepsilon_{\mathrm{sep}}(t)=\norm{\mathbf{u}^\star(t)-\qmap_{t}(\mathbf{x}_{\mathrm{EMT}}(t))}$ is the distance between the ideal effective state and the solution $\mathbf{x}_{\mathrm{EMT}}$ of~\eqref{eq:emt} in the transformed coordinates, so the error with respect to full EMT simulation at $T$ is bounded by~\eqref{eq:global} plus $\varepsilon_{\mathrm{sep}}(T)$.

\subsection{Computational cost and speedup}\label{sec:cost}
The same steps that govern accuracy determine the cost. Let $c_\mu$ be the cost of a microscopic step, $c_{M}$ the additional cost of an inner cycle, and $c_{Q}$ the cost of the outer convolution, Jacobian, filter, and propagation. \rev{For constant steps and windows,} an outer cycle contains $\eta_{Q}/\Delta_{M}$ inner cycles of $\eta_\mu/h_\mu$ microscopic steps, so the cost per unit simulated time is
\begin{equation}\mathcal C_3\approx\frac{1}{\eta_{Q}+h_{Q}}\left[\frac{\eta_{Q}}{\eta_\mu+h_{M}}\Bigl(\frac{\eta_\mu}{h_\mu}c_\mu+c_{M}\Bigr)+c_{Q}\right],\label{eq:cost}\end{equation}
to which the detailed intervals after events and the deferred outer attempts add their cost. When microscopic integration dominates, the speedup over full EMT simulation, whose cost is $c_\mu/h_\mu$, is
\begin{equation}S_3\approx\Bigl(1+\frac{h_{M}}{\eta_\mu}\Bigr)\Bigl(1+\frac{h_{Q}}{\eta_{Q}}\Bigr).\label{eq:speedup}\end{equation}
The first factor is the gain of the two-level HMM, and the second is the gain of the outer level, which multiplies it. Since $h_{Q}/\eta_{Q}$ reaches tens to more than one hundred during quasi-steady evolution, the outer level dominates the speedup, provided that $c_{Q}$ stays small.

The steps and windows in~\eqref{eq:speedup} are bounded by the accuracy requirement through Theorem~\ref{thm:nested}, which gives the speedup attainable at a prescribed accuracy.
\begin{corollary}[Accuracy-constrained speedup]\label{cor:speedup}
For a tolerance $\vartheta$ and constants in~\eqref{eq:averaging} and~\eqref{eq:global} independent of $\epsilon_{f,\ell}$, every term of~\eqref{eq:global} is at most $C\vartheta$ when $h_\mu^r$, $h_{M}$, $h_{Q}^3$, and $\varrho$ are at most $\vartheta$ and the windows satisfy $\epsilon_{f,\ell}\vartheta^{-1/k_\ell}\le\eta_\ell\le\vartheta^{1/p_\ell}$, $\ell=M,Q$, up to constants. With the shortest such windows and the longest such outer step, which is $h_{Q}=\vartheta^{1/3}$ when $\varrho$ restricts neither the step nor the windows,~\eqref{eq:speedup} becomes
\begin{equation}S_3\sim\Bigl(1+\frac{h_{M}\,\vartheta^{1/k_{M}}}{\epsilon_\mu}\Bigr)\Bigl(1+\frac{\vartheta^{1/3+1/k}}{\epsilon_{M}}\Bigr),\label{eq:speedlaw}\end{equation}
and the microscopic integration costs $1/(h_\mu S_3)$ steps per unit simulated time.
\end{corollary}
The first factor is the gain of the two-level HMM~\cite{huanghmm}. \rev{The second is the gain of the outer level. Its window is set by the electromechanical scale $\epsilon_{M}$, and its step by the quasi-steady integration error, since Propositions~\ref{prop:stability} and~\ref{prop:sensitivity} bound the effect of residual fast components independently of $h_{Q}$.} Since $h_{M}$ resolves the electromechanical dynamics, it is at most of order $\epsilon_{M}$, so the two-level factor is at most of order $\epsilon_{M}/\epsilon_\mu$, whereas with $h_{M}$ of order $\epsilon_{M}$ the product in~\eqref{eq:speedlaw} grows as $1/\epsilon_\mu$, the full separation between the electromagnetic and quasi-steady scales. \rev{The window interval is nonempty when $\epsilon_{f,\ell}\lesssim\vartheta^{1/p_\ell+1/k_\ell}$, which requires $\epsilon_{M}\lesssim\vartheta^{1+1/k}$ for a symmetric outer kernel ($p=1$) but only $\epsilon_{M}\lesssim\vartheta^{1/3+1/k}$ for $p=3$, consistent with the drift of the symmetric-kernel variant in Fig.~\ref{fig:ablation}.} 
\section{Case studies}\label{sec:cases}
\subsection{Test systems and evaluation}\label{sec:settings}
\begin{figure}[!ht]
\color{black}
\centering\includegraphics[width=.7\columnwidth]{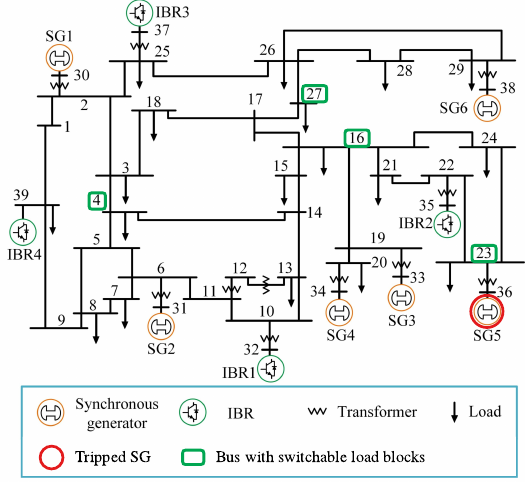}
\caption{Modified IEEE 39-bus system.}\label{fig:network}
\end{figure}

The test system in Fig.~\ref{fig:network} is the modified IEEE 39-bus EMT model of~\cite{huanghmm}. Six SGs use voltage-behind-reactance models with field and damper windings, SEXS excitation, and TGOV1 turbine--governor control~\cite{huanghmm,wang}. Four grid-following IBRs include PLL, power and voltage regulation, current control, converter-voltage dynamics, and interface inductances, and all devices connect through a three-phase dynamic network with RL loads. Four switchable RL load blocks, secondary frequency regulation, and a restoration supervisor are added for the long-term studies, which gives $N_x=430$. The compression map uses the rotor angle of SG1, or of SG2 when SG1 is disconnected, as the area reference.

Two fixed-step EMT solvers serve as micro-solvers, fourth-order Runge--Kutta (RK4) with a step of $10~\upmu$s and the nodal trapezoidal (TR) solver of EMT programs~\cite{dommel,lincda}, detailed in Appendix~\ref{app:impl}, with a step of $5~\upmu$s. Each comparison uses the same micro-solver for all methods, and two-level HMM shares the maps, the middle update, and the transitions with three-level HMM, so the two differ only by the outer level. The settings S1--S4 in Table~\ref{tab:settings} differ in the middle step cap and tolerance and in $\sigma_{Q}$. Two-level HMM uses the middle settings of S2 and S4, and S2 is used unless stated otherwise.

\begin{table}[!t]
\caption{HMM Settings}\label{tab:settings}
\centering\footnotesize
\setlength{\tabcolsep}{4pt}
\begin{tabular}{@{}ll@{}}\toprule
Quantity & Value\\\midrule
Middle window $\eta_\mu$, hand-over & 26.4 ms, $e_\mu\le1$ in two windows\\
Outer $\eta_{Q}$, $w$, orders $p$ & 0.5 s, 0.25 s, 3 and 2\\
Filter $\tau$, minimum step $h_{Q,\min}$ & 0.5 s, 1 s\\
Scales of angles, frequencies, others & 0.02 rad, 2 mHz, $0.002(1+|u_i|)$\\
Scale of secondary command & $5.05\times10^{-4}$ pu (0.05 MW)\\\midrule
Middle cap, tolerance, $\sigma_{Q}$ & \\
\quad S1, S2, S3 & 39.6 ms, 0.01, 1, 1/4, 1/16\\
\quad S4 & 19.8 ms, 0.005, 1/100\\\bottomrule
\end{tabular}
\end{table}

The generator angle error is the maximum error of the rotor angles $\theta_i$ of the connected SGs, the relative angle error refers to $\theta_i-\theta_1$, or $\theta_i-\theta_2$ when SG1 is disconnected, and the SG frequency is $f_0(1+\Delta\omega_i)$ with $f_0=60$~Hz. These errors are maxima over all outputs and the interior points of the outer steps. Waveforms are compared at matching physical times without phase alignment by the normalized root-mean-square error $\mathrm{NRMSE}=100\,(\sum_j\|\mathbf{y}(t_j)-\mathbf{y}_{\mathrm{EMT}}(t_j)\|_2^2/\sum_j\|\mathbf{y}_{\mathrm{EMT}}(t_j)\|_2^2)^{1/2}\,\%$ for a three-phase current $\mathbf{y}$ in per unit (pu) over a 100-ms window, which in the restoration starts at each pickup. All simulations ran on an Apple M4 Pro in serial processes with one numerical-library thread.

\subsection{Restoration over 500 s}\label{sec:restoration}
The restoration study follows a generation loss, minutes of frequency regulation, and four load pickups along one trajectory. SG5 trips at 0.1~s, four 60-MW load blocks at buses 4, 16, 23, and 27 are shed at 0.3~s, and from 10~s secondary regulation integrates the filtered center-of-inertia (COI) frequency error into a command, rate-limited to 1~MW/s, for the governors of the connected SGs, whose scale in Table~\ref{tab:settings} is the speed scale multiplied by the sum of their inverse droops. From 60~s, a supervisor checks frequency, voltage, and generation-margin conditions every 10~s and reconnects the next block after two successive passes, at least 40~s after the previous pickup, each method evaluating the conditions from its own state.

\begin{figure}[!ht]
\color{black}
\centering
\includegraphics[width=\columnwidth]{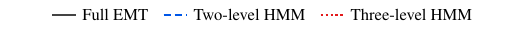}\\[1pt]
\subfloat[]{\includegraphics[width=\columnwidth]{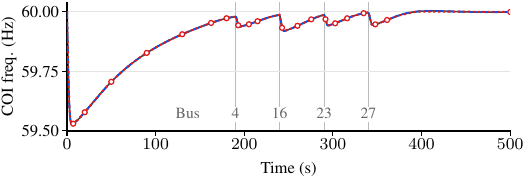}%
\label{fig:4a}}\\
\subfloat[]{\includegraphics[width=\columnwidth]{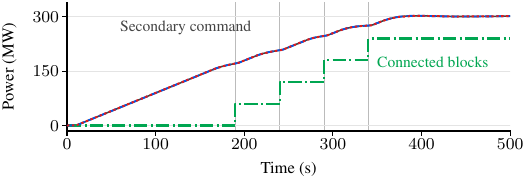}%
\label{fig:4b}}\\
\subfloat[]{\includegraphics[width=\columnwidth]{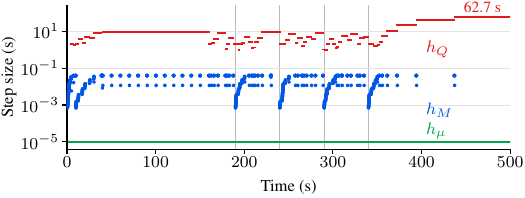}%
\label{fig:4c}}
\caption{Restoration over 500 s with RK4 and S2. (a) COI frequency, with circles at selected outer endpoints. (b) Secondary command and connected switchable demand. (c) Step sizes of three-level HMM. Vertical lines mark the load pickups.}\label{fig:overview}
\end{figure}
\begin{figure}[!ht]
\color{black}
\centering
\includegraphics[width=\columnwidth]{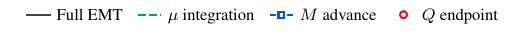}\\[-6pt]
\subfloat[]{\includegraphics[width=.49\columnwidth]{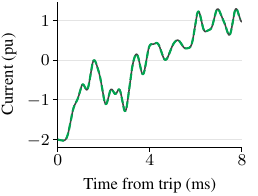}%
\label{fig:5a}}
\hfil
\subfloat[]{\includegraphics[width=.49\columnwidth]{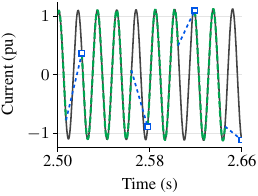}%
\label{fig:5b}}\\
\subfloat[]{\includegraphics[width=.49\columnwidth]{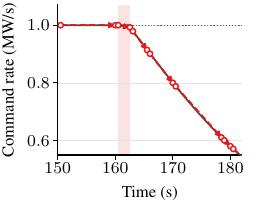}%
\label{fig:5c}}
\hfil
\subfloat[]{\includegraphics[width=.49\columnwidth]{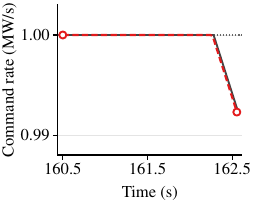}%
\label{fig:5d}}\\
\subfloat[]{\includegraphics[width=\columnwidth]{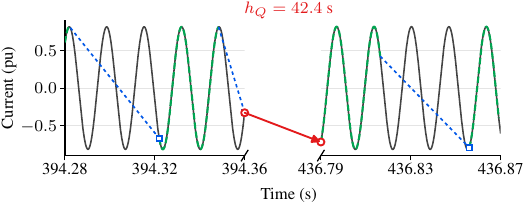}%
\label{fig:5e}}
\caption{Local behavior in the restoration study. (a) Branch 17--18 phase-a current after the SG5 trip. (b) The same current during $\mu/M$ cycles. (c) Rate of the secondary command around the release of its rate limit. (d) Zoom on the shaded outer step, with the rate within the step from~\eqref{eq:stage}--\eqref{eq:qstep} (dashed). (e) Branch 17--18 phase-a current around a long outer step.}\label{fig:local}
\end{figure}

Fig.~\ref{fig:overview} shows that the three levels follow the complete process. With either micro-solver, full EMT simulation and both HMM methods reconnect the blocks at 190, 240, 290, and 340~s and follow the regulation with indistinguishable COI frequencies. Detailed integration resolves the transient of every switching event, as for the SG5 trip in Fig.~\ref{fig:local}(a). The transition indicator then hands the simulation to $\mu/M$ cycles, which follow the electromechanical response in Fig.~\ref{fig:local}(b), and outer steps take over once the response becomes quasi-steady. Between pickups, the supervisory checks bound the outer steps, and after the last pickup the step grows to 62.7~s, so that outer steps cover 90\% of the horizon.

\begin{table}[!t]
\caption{500-s Restoration (Maximum Errors)}\label{tab:results}
\centering\footnotesize
\setlength{\tabcolsep}{2pt}
\begin{tabular*}{\columnwidth}{@{}l@{\extracolsep{\fill}}lrrrrrr@{}}\toprule
 & & Time & Speedup & Gen.\ angle & Rel.\ angle & SG freq. & NRMSE$^{\mathrm{a}}$\\
\multicolumn{2}{@{}l}{Method} & (s) & & ($^\circ$) & ($^\circ$) & (mHz) & (\%)\\\midrule
\multicolumn{2}{@{}l}{RK4} & & & & & & \\
\quad Full EMT & & 221.40 & 1.00 & --- & --- & --- & ---\\
\quad Two & S2 & 96.90 & 2.28 & 0.821 & 0.058 & 0.859 & 1.24\\
\quad Three & S2 & 13.51 & 16.38 & 0.568 & 0.056 & 0.834 & 0.96\\
\quad Two & S4 & 130.94 & 1.69 & 0.278 & 0.018 & 0.286 & 0.25\\
\quad Three & S4 & 25.10 & 8.82 & 0.248 & 0.017 & 0.273 & 0.23\\\midrule
\multicolumn{2}{@{}l}{TR} & & & & & & \\
\quad Full EMT & & 428.07 & 1.00 & 0.074 & 0.000 & 0.001 & 0.12\\
\quad Two & S2 & 189.75 & 2.26 & 2.489 & 0.058 & 0.860 & 4.13\\
\quad Three & S2 & 24.36 & 17.57 & 2.125 & 0.056 & 0.834 & 3.54\\\bottomrule
\end{tabular*}
\par\smallskip\begin{minipage}{\columnwidth}\footnotesize
$^{\mathrm{a}}$After the pickups. Speedups refer to full EMT simulation with the same solver, and errors to full EMT simulation with RK4.
\end{minipage}
\end{table}

The outer level also advances across nonsmooth limiter switching. The secondary command rises at its rate limit until the filtered COI frequency exceeds 59.95~Hz, and the outer steps in Fig.~\ref{fig:local}(c) pass this release. In the zoom of Fig.~\ref{fig:local}(d), the rate within the step aligns with full EMT simulation on both sides of the switching, because the nonlinear correction in~\eqref{eq:qstep} evaluates the EMT field, and with it the limiter, at the propagated state. \revb{Fig.~\ref{fig:local}(c)--(d) thus provides empirical evidence that the outer step can remain accurate across this limiter release, even though this case is not covered by Theorem~\ref{thm:nested}.} After the long outer step in Fig.~\ref{fig:local}(e), the microscopic window started from the reconstructed EMT state coincides with the full EMT current. Table~\ref{tab:results} compares accuracy and cost. With S2, three-level HMM is 16.4 and 17.6 times faster than full EMT simulation with RK4 and TR, and 7.2 and 7.8 times faster than two-level HMM, with lower errors in every quantity. With S4 and RK4, both methods reach generator angle errors near $0.25^\circ$ and post-pickup NRMSEs of at most 0.25\%, and three-level HMM is 5.2 times faster. As~\eqref{eq:speedup} predicts, the speedup follows the simulated time covered by microscopic integration, 27.0~s of the 500~s with S2 against 214.6~s for two-level HMM. With TR, the generator angles of both HMM methods deviate more from the RK4 reference while their relative angles and frequencies are unchanged, so this deviation arises in the $\mu/M$ cycles that both methods share.

\begin{figure}[!t]
\color{black}
\centering
\subfloat[]{\includegraphics[width=.49\columnwidth]{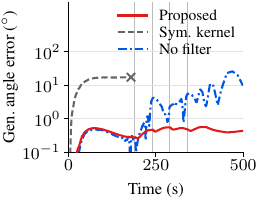}%
\label{fig:7a}}
\hfil
\subfloat[]{\includegraphics[width=.49\columnwidth]{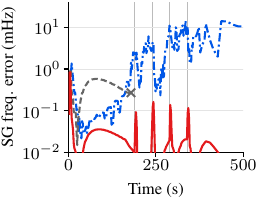}%
\label{fig:7b}}
\caption{Outer-level variants in the restoration with RK4 and S2, as 2-s maxima. (a) Generator angle error. (b) SG frequency error. Crosses mark the first pickup of the symmetric-kernel variant.}\label{fig:ablation}
\end{figure}

\begin{figure}[!t]
\color{black}
\centering\includegraphics[width=\columnwidth]{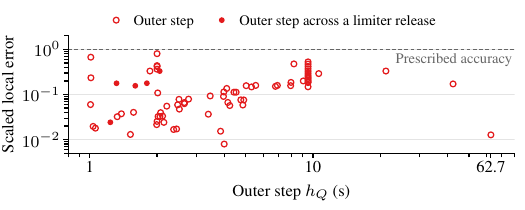}
\caption{\revb{Local error of every accepted outer step in the restoration with RK4 and S2, from full EMT simulation restarted at the start of the step.}}\label{fig:localerr}
\end{figure}

Fig.~\ref{fig:ablation} isolates the two designs of the outer level, a symmetric kernel over the outer history and an outer step without the resolvent filter. Both variants replay the outer-step times of the proposed method and advance their own states. With the symmetric kernel, the force describes the center of the history rather than its end, so the generator angles drift by up to $17^\circ$ during the regulation and the supervisor reconnects every block 10~s earlier than in full EMT simulation. Without the resolvent filter, the residual fast content of the convolution force is propagated over the long outer steps, so the errors grow after the pickups to $26^\circ$ within the longest steps. Only the combination of both designs reproduces the regulation and the restoration sequence of full EMT simulation. \revb{The local error of each outer step is shown in Fig.~\ref{fig:localerr}. All 77 outer steps stay within the prescribed accuracy, including the five steps that cross a release of the command rate limit, with local errors of at most 0.33. The largest local errors, up to 0.80, occur in the short steps that follow a pickup, while the 21.2-, 42.4-, and 62.7-s steps reach 0.33, 0.17, and 0.013, so the long outer steps remain accurate, in line with the step-independent bounds of Propositions~\ref{prop:stability} and~\ref{prop:sensitivity}.}

\subsection{Accuracy and cost across settings and disturbances}\label{sec:tradeoff}
In the outage study, nine simulations disconnect SG1--SG6 or IBR1--IBR3 at 0.1~s from the same operating point and continue to 100~s with secondary regulation active from the start. Its rate-limited command raises generation over the whole horizon, so each case covers the primary response and the secondary regulation that follows.

\begin{figure}[!ht]
\color{black}
\centering\includegraphics[width=\columnwidth]{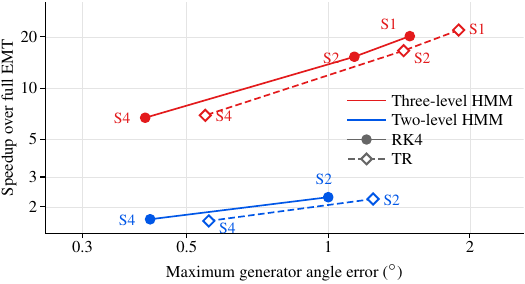}
\caption{Speedup versus generator angle error for the nine source outages, averaged as in Table~\ref{tab:n1}.}\label{fig:tradeoff}
\end{figure}
\begin{table}[!t]
\caption{Nine Source Outages (Means over the Cases)}\label{tab:n1}
\centering\footnotesize
\setlength{\tabcolsep}{2pt}
\begin{tabular*}{\columnwidth}{@{}l@{\extracolsep{\fill}}lrrrrr@{}}\toprule
 & & Speedup$^{\mathrm{a}}$ & Gen.\ angle & Rel.\ angle & SG freq. & NRMSE\\
\multicolumn{2}{@{}l}{Method} & & ($^\circ$) & ($^\circ$) & (mHz) & (\%)\\\midrule
\multicolumn{2}{@{}l}{RK4} & & & & & \\
\quad Two & S2 & 2.28 & 1.00 & 0.079 & 1.49 & 1.52\\
\quad Three & S2 & 15.29 & 1.14 & 0.071 & 1.36 & 1.77\\
\quad Two & S4 & 1.69 & 0.42 & 0.023 & 0.45 & 0.37\\
\quad Three & S4 & 6.69 & 0.41 & 0.022 & 0.44 & 0.36\\\midrule
\multicolumn{2}{@{}l}{TR} & & & & & \\
\quad Two & S2 & 2.23 & 1.25 & 0.079 & 1.49 & 2.10\\
\quad Three & S2 & 16.61 & 1.45 & 0.071 & 1.36 & 2.45\\
\quad Two & S4 & 1.65 & 0.56 & 0.023 & 0.45 & 0.90\\
\quad Three & S4 & 6.92 & 0.55 & 0.022 & 0.44 & 0.88\\\bottomrule
\end{tabular*}
\par\smallskip\begin{minipage}{\columnwidth}\footnotesize
$^{\mathrm{a}}$Geometric mean. Errors refer to full EMT simulation with RK4, and NRMSE to the window after the largest outer step.
\end{minipage}
\end{table}

Fig.~\ref{fig:tradeoff} relates accuracy and speedup for these outages. With either micro-solver, the three-level runs lie above the two-level runs at all tested settings, and a smaller $\sigma_{Q}$ trades speedup for accuracy until the error levels off, consistent with the middle-level contribution in Theorem~\ref{thm:nested}, which S4 lowers through its smaller middle step cap and tolerance. In the restoration, the generator angle error falls from $2.17^\circ$ with S1 to $0.57^\circ$ with S2, remains at this level with S3 ($0.86^\circ$), and decreases to $0.25^\circ$ only with S4. Table~\ref{tab:n1} shows that, with S4, three-level HMM matches the errors of two-level HMM while it is 4.0 and 4.2 times faster with RK4 and TR. With S2, it is 15.3 and 16.6 times faster than full EMT simulation. The three-level speedup is larger with TR, because TR costs more per simulated second while the outer-level overhead stays the same.

\subsection{390-bus system}\label{sec:large}
Ten copies of the 39-bus model, without secondary regulation and supervisor, are joined by three-phase RL tie lines between bus~39 of every pair of areas, which gives $N_x=4435$. A phase-A-to-ground fault at bus~16 of area~1 is applied at 0.1~s and cleared after five cycles, and the system is simulated for 50~s with S2 and the nodal TR solver at $10~\upmu$s.
\begin{table}[!t]
\caption{Component Costs for the 390-Bus System}\label{tab:sparse}
\centering\footnotesize
\begin{tabular}{@{}lrr@{}}\toprule
Component & Sparse & Dense\\\midrule
TR step (incl.\ nodal solve) & 53.2 $\upmu$s & 1015.8 $\upmu$s\\
Nodal solve & 13.6 $\upmu$s & 490.3 $\upmu$s\\
Factorization of $\mathbf{I}-\tau\mathbf{J}$ & 0.0047 s & 1.16 s\\
$\varphi$-function actions of one step & 0.068 s & 203 s\\
Full EMT simulation, 50 s & 270.3 s & 5083 s\\\bottomrule
\end{tabular}
\end{table}
\begin{figure}[!t]
\color{black}
\centering
\subfloat[]{\includegraphics[width=.49\columnwidth]{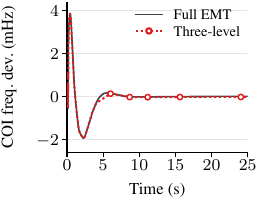}%
\label{fig:8a}}
\hfil
\subfloat[]{\includegraphics[width=.49\columnwidth]{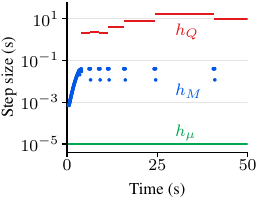}%
\label{fig:8b}}\\
\includegraphics[width=\columnwidth]{fig5_legend.pdf}\\[1pt]
\subfloat[]{\includegraphics[width=\columnwidth]{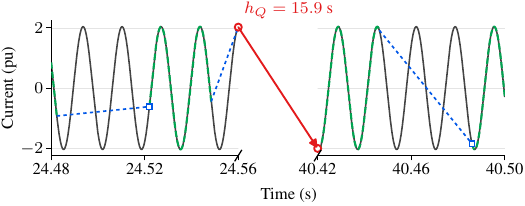}%
\label{fig:8c}}
\caption{390-bus system after the fault in area~1. (a) COI frequency deviation. (b) Step sizes of three-level HMM\rev{, with the fixed TR step $h_\mu$}. (c) Branch 17--18 phase-a current in area~1 around the longest outer step.}\label{fig:large}
\end{figure}
Three-level HMM completes the simulation 11.0 times faster than full EMT simulation with the same sparse solver, with outer steps of up to 15.9~s (Fig.~\ref{fig:large}(b)), a generator angle error of $0.57^\circ$, and an SG frequency error of 1.2~mHz. After the longest outer step, the microscopic window started from the reconstructed EMT state coincides with the full EMT current in Fig.~\ref{fig:large}(c), with an NRMSE of 0.68\%. The sparse realization in Table~\ref{tab:sparse} makes this possible: the sparse nodal factorization makes each TR step 19 times cheaper, the $\varphi$-function actions of an outer step take 0.068~s, and the outer computations take only 1.1~s of the 24.5-s run, in line with~\eqref{eq:cost}.

\section{Conclusion}\label{sec:conclusion}
\rev{This paper has built a nested multiscale framework that adaptively simulates electromagnetic, electromechanical, and quasi-steady-state dynamics on one unified EMT model with an unmodified fixed-step solver. The analysis establishes the stability of long outer steps and shows that the outer level multiplies the speedup of the two-level HMM. In regulation, restoration, and outage studies and on a 390-bus system, the framework reproduced full EMT simulation through every switching transient, more than ten times faster at the default setting and several times faster than two-level HMM at matched accuracy. Because the solver enters only through state-to-state advances, the framework applies to EMT models and solvers with other device, control, and network representations.}
\appendices
\section{Proofs}\label{app:proofs}
\begin{IEEEproof}[Proof of Lemma~\ref{lem:transfer}]
Subtracting the second forms of~\eqref{eq:conv} and applying the integral norm inequality gives the bound with $\int_{-\eta}^0|K'_\eta(s)|\,ds=\kappa_K/\eta$.
\end{IEEEproof}
\begin{IEEEproof}[Proof of Lemma~\ref{lem:bias}]
Taylor expansion of $\mathbf{v}(t+s)$ about $t$ with integral remainder of order $p$ and the moment conditions~\eqref{eq:moments} remove all terms of degree below $p$, and bounding the remainder gives~\eqref{eq:bias}.
\end{IEEEproof}
\begin{IEEEproof}[Proof of Proposition~\ref{prop:stability}]
Exact sampling gives $u_{M}(\eta_{Q}+s)=e^{\lambda(\eta_{Q}+s)}u_0$ and endpoint $u=e^{\lambda\eta_{Q}}u_0$. The first form of~\eqref{eq:conv} gives $F_{Q}=\lambda(D_K+e^{\lambda\eta_{Q}})u_0$, and since $g=\lambda u$, \eqref{eq:resolvent} yields $\widetilde F_{Q}=\lambda e^{\lambda\eta_{Q}}u_0+\lambda D_Ku_0/(1-\tau\lambda)^2$. The field is linear, so $N=0$, and~\eqref{eq:stage}--\eqref{eq:qstep} give $u^+=u+(e^{\lambda h_{Q}}-1)\widetilde F_{Q}/\lambda$, which is~\eqref{eq:amplification}. Since $|e^{\lambda(\eta_{Q}+h_{Q})}|\le e^{\eta_{Q}\operatorname{Re}\lambda}$ and $|e^{\lambda h_{Q}}-1|\le2$, the first inequality in~\eqref{eq:rhobar} follows, and the second follows from $|e^{\lambda(\eta_{Q}+s)}|\le1$ on the support of $K_{Q}$.
\end{IEEEproof}
\begin{IEEEproof}[Proof of Proposition~\ref{prop:sensitivity}]
The perturbation enters $\mathbf{u}^+$ through $h_{Q}\varphi_1(h_{Q}\mathbf{J})\mathbf{S}$, whose modal value is $G$, and $|(e^{\lambda h}-1)/\lambda|=|\int_0^he^{\lambda\xi}d\xi|\le\min\{h,|\operatorname{Re}\lambda|^{-1}\}$. For the affine trajectory, differentiating $\mathbf{g}(\mathbf{u}_*+t\mathbf{c})=\mathbf{c}$ gives $\mathbf{J}\mathbf{c}=\mathbf{0}$. A constant-reproducing kernel returns $\mathbf{F}_{Q}=\mathbf{c}=\mathbf{g}$, hence $\widetilde{\mathbf{F}}_{Q}=\mathbf{c}$, $\mathbf{d}=h_{Q}\varphi_1(h_{Q}\mathbf{J})\mathbf{c}=h_{Q}\mathbf{c}$, and $\mathbf{N}(\mathbf{d})=\mathbf{0}$ because $\mathbf{g}$ is constant along the line.
\end{IEEEproof}
\begin{IEEEproof}[Proof of Theorem~\ref{thm:nested}]
Let $e_n$ be the error at the start of outer cycle $n$, $\Delta_{Q,n}=\eta_{Q}+h_{Q,n}$, and $\Phi_n$ the ideal effective cycle, the flow of $\mathbf{F}_{M}^\star$ over $\eta_{Q}$ followed by that of $\mathbf{F}_{Q}^\star$ over $h_{Q,n}$, with Lipschitz constant $e^{C\Delta_{Q,n}}$. Then $e_{n+1}\le e^{C\Delta_{Q,n}}e_n+\mathcal D_n$, with $\mathcal D_n$ the distance between the numerical cycle and $\Phi_n$ from the same state $\mathbf{u}_n$, and the discrete Gronwall inequality over $\sum_n\Delta_{Q,n}\le T$ gives~\eqref{eq:global} once $\mathcal D_n$ is bounded by $C\Delta_{Q,n}$ times the bracket in~\eqref{eq:global}. Within the sampling of cycle $n$, each $\mu/M$ cycle is compared with the flow of $\mathbf{F}_{M}^\star$ over $\Delta_{M}$ from the start of its micro window. The micro solution stays within $C\eta_\mu h_\mu^r+\eta_\mu\varrho$ of the flow of $\mathbf{g}$, the second term being the start-up error, so Lemma~\ref{lem:transfer} and~\eqref{eq:averaging} give a force~\eqref{eq:mforce} within $C\delta_{M}$ of $\mathbf{F}_{M}^\star$ at the end of this flow, with $\delta_{M}=h_\mu^r+\eta_\mu^{p_{M}}+(\epsilon_\mu/\eta_\mu)^{k_{M}}+\varrho$. This end lies within $\Delta_{M}\varrho$ of the flow of $\mathbf{F}_{M}^\star$, and the flows of $\mathbf{F}_{M}^\star$ over $h_{M}$ from the two ends differ by at most $e^{Ch_{M}}\Delta_{M}\varrho$. With the local error $Ch_{M}^2\le C\Delta_{M}h_{M}$ of~\eqref{eq:madvance} from this end, the cycle ends within $C\Delta_{M}\delta_{M}$ of the flow of $\mathbf{F}_{M}^\star$ over $\Delta_{M}$, with $h_{M}$ added to $\delta_{M}$ from here on. Propagating these errors with the Lipschitz constant $e^{C\Delta_{M}}$ of the flow bounds the distance of the middle-level history from the flow of $\mathbf{F}_{M}^\star$ from $\mathbf{u}_n$ by $E_{M}\le C\eta_{Q}\delta_{M}$, so by Lemma~\ref{lem:transfer} and~\eqref{eq:averaging} the outer force satisfies $\norm{\mathbf{F}_{Q}-\mathbf{F}_{Q}^\star}\le C(\delta_{M}+\eta_{Q}^p+(\epsilon_{M}/\eta_{Q})^k)$, with $\mathbf{F}_{Q}^\star$ at the exact-history endpoint. Evaluated at this endpoint, \eqref{eq:resolvent} gives a force that differs from $\mathbf{F}_{Q}^\star$ by $\mathbf{S}(\mathbf{F}_{Q}-\mathbf{F}_{Q}^\star)$ plus the filter defect. The endpoint error $E_{M}$ passes through the outer step, which evaluates~\eqref{eq:resolvent} at the computed endpoint, with constant $C$, the force error with constant $Ch_{Q}$, the local error of the third-order step is at most $Ch_{Q}^4$, and the filter defect, which perturbs $\mathbf{d}$ and passes through~\eqref{eq:qstep} with constant $C$, the additional local error of the outer step, and the matrix-function and Jacobian errors, all part of $\varrho$, add $C\Delta_{Q,n}\varrho$. Summing these contributions with $\eta_{Q},h_{Q}\le\Delta_{Q,n}$ gives the required bound on $\mathcal D_n$. At a common switching time, the state-reset map $\mathcal T_E$ is applied to both states, so $e^+\le L_Ee^-$ with its Lipschitz constant $L_E$. Intervals of detailed integration and of $\mu/M$ cycles without outer steps are covered by the middle-level estimate above, with the exact EMT solution and the trajectory of $\mathbf{F}_{M}^\star$ as references, and a deferred outer attempt leaves the state unchanged, so the reset bound and the Gronwall argument extend~\eqref{eq:global} over $[0,T]$.
\end{IEEEproof}
\begin{IEEEproof}[Proof of Corollary~\ref{cor:speedup}]
Inside the window interval both terms of~\eqref{eq:averaging} are at most $C\vartheta$, and inserting the lower ends of the intervals and $h_{Q}=\vartheta^{1/3}$ into~\eqref{eq:speedup} gives~\eqref{eq:speedlaw}.
\end{IEEEproof}
\section{Kernels and numerical implementation}\label{app:impl}
\rev{The smooth kernel in~\eqref{eq:endkernel} is $b_w(s)=c_w\exp(-\nu/(1-\zeta^2))$ with $\zeta=1+2s/w$ on $(-w,0)$, a shape parameter $\nu>0$, and $c_w$ normalizing its integral, and the convolutions use $\mathbf{u}(t+s)-\mathbf{u}(t)$ in place of $\mathbf{u}(t+s)$, which leaves~\eqref{eq:conv} unchanged and avoids cancellation for angle states. With $\mathbf{J}_{\mathcal Q}=D_{\mathbf{x}}\qmap_t(\mathbf{x})$ and $\mathbf{J}_x=D_{\mathbf{x}}\mathbf{f}_\sigma(\mathbf{x},t)$, the variational equations give $\mathbf{J}=(\dot{\mathbf{J}}_{\mathcal Q}+\mathbf{J}_{\mathcal Q}\mathbf{J}_x)\mathbf{J}_{\mathcal Q}^{-1}$, where $\dot{\mathbf{J}}_{\mathcal Q}$ contains the variation of the rotation angles, and $\mathbf{J}$ is evaluated by central differences of~\eqref{eq:transformed}, perturbing together the local columns of areas that share no tie line. One sparse factorization of $\mathbf{I}-\tau\mathbf{J}$ serves the filter~\eqref{eq:resolvent} and shift-and-invert Arnoldi approximations of the $\varphi$-function actions~\cite{niesen,botchev}. The nodal trapezoidal solver uses companion models of the inductive branches and nodal capacitances~\cite{dommel}, factorizes the conductance matrix once per step size and topology after a block minimum-degree ordering~\cite{tinney}, integrates the devices by a trapezoidal predictor--corrector, and applies the critical damping adjustment~\cite{lincda} at events and at the start of each microscopic window.}
\section*{Acknowledgment}
The original manuscript was drafted by the authors. Claude (Anthropic) was used to assist with language editing and polishing in Sections I–V and the appendices. The authors verified all content and take full responsibility for the manuscript.
\bibliographystyle{IEEEtran}
\bibliography{Ref}
\end{document}